\documentclass[floatfix,twocolumn,aps,prl,reprint,superscriptaddress]{revtex4-1}

\usepackage{amsmath}
\usepackage{amssymb}
\usepackage{float}
\usepackage{graphicx}
\usepackage{color,soul}
\usepackage{hyperref}
\usepackage[caption=false]{subfig}
\usepackage[colorinlistoftodos]{todonotes}
\usepackage{lipsum}

\usepackage{tikz}
\usepackage{circuitikz}
\usepackage{tikz-timing}
\usetikzlibrary{arrows.meta,decorations.pathmorphing,decorations.pathreplacing,positioning,shapes,fadings}

\graphicspath{ {main_figures/} {supplement_figures/} }

\makeatletter
\@ifundefined{textcolor}{}
{%
	\definecolor{BLACK}{gray}{0}
	\definecolor{WHITE}{gray}{1}
	\definecolor{RED}{rgb}{1,0,0}
	\definecolor{GREEN}{rgb}{0,1,0}
	\definecolor{BLUE}{rgb}{0,0,1}
	\definecolor{CYAN}{cmyk}{1,0,0,0}
	\definecolor{MAGENTA}{cmyk}{0,1,0,0}
	\definecolor{YELLOW}{cmyk}{0,0,1,0}
}
\usepackage{dcolumn}
\usepackage{bm}
\usepackage{float}
\makeatother

\begin{document}

\title{Atomic layer deposition of titanium nitride for quantum circuits}

\author{Abigail Shearrow}
\affiliation{James Franck Institute, University of Chicago, Chicago, Illinois 60637, USA}
\affiliation{Department of Physics, University of Chicago, Chicago, Illinois 60637, USA}

\author{Gerwin Koolstra}
\affiliation{James Franck Institute, University of Chicago, Chicago, Illinois 60637, USA}
\affiliation{Department of Physics, University of Chicago, Chicago, Illinois 60637, USA}

\author{Samuel J. Whiteley}
\affiliation{Department of Physics, University of Chicago, Chicago, Illinois 60637, USA}
\affiliation{Institute for Molecular Engineering, University of Chicago, Chicago, Illinois 60637, USA}

\author{Nathan Earnest}
\affiliation{James Franck Institute, University of Chicago, Chicago, Illinois 60637, USA}
\affiliation{Department of Physics, University of Chicago, Chicago, Illinois 60637, USA}

\author{Peter S. Barry}
\affiliation{Department of Astronomy and Astrophysics, University of Chicago, Chicago, Illinois 60637, USA}

\author{F. Joseph Heremans}
\affiliation{Institute for Molecular Engineering and Materials Science Division, Argonne National Laboratory, Lemont, Illinois 60439, USA}

\author{David D. Awschalom}
\affiliation{Institute for Molecular Engineering, University of Chicago, Chicago, Illinois 60637, USA}
\affiliation{Institute for Molecular Engineering and Materials Science Division, Argonne National Laboratory, Lemont, Illinois 60439, USA}

\author{Erik Shirokoff}
\affiliation{Department of Astronomy and Astrophysics, University of Chicago, Chicago, Illinois 60637, USA}

\author{David I. Schuster}
\email{David.Schuster@uchicago.edu}
\affiliation{James Franck Institute, University of Chicago, Chicago, Illinois 60637, USA}
\affiliation{Department of Physics, University of Chicago, Chicago, Illinois 60637, USA}
\affiliation{Institute for Molecular Engineering and Materials Science Division, Argonne National Laboratory, Lemont, Illinois 60439, USA}

\date{\today }

\begin{abstract}
	Superconducting thin films with high intrinsic kinetic inductance are of great importance for photon detectors, achieving strong coupling in hybrid systems, and protected qubits. We report on the performance of titanium nitride resonators, patterned on thin films (9-110 nm) grown by atomic layer deposition, with sheet inductances of up to 234 pH/$\square$. For films thicker than 14 nm, quality factors measured in the quantum regime range from $0.4$ to $1.0$ million and are likely limited by dielectric two-level systems. Additionally, we show characteristic impedances up to 28 k$\Omega$, with no significant degradation of the internal quality factor as the impedance increases. These high impedances correspond to an increased single photon coupling strength of 24 times compared to a 50 $\Omega$ resonator, transformative for hybrid quantum systems and quantum sensing.
\end{abstract}

\maketitle

In a superconductor, kinetic inductance (KI) is the inductance due to the inertia of cooper pairs. Films with large KI are attractive for a wide range of superconducting device applications, including high frequency single photon detectors \cite{DayNature2003}. In addition, the degree of non-linearity of the KI with applied DC current opens up the possibility of novel devices such as superconducting phase shifters \cite{CheArxiv2017}, ultra-sensitive current sensors \cite{Kher2016} and quantum-limited travelling-wave parametric amplifiers \cite{EomNatPhys2012}. Recently, high KI materials have also shown great promise in hybrid quantum systems that aim to couple microwave photons to spin degrees of freedom \cite{Samkharadze2018Science, LandigArxiv2017}. These materials offer increased spin-photon coupling due to larger zero-point fluctuations of the electric field \cite{SamkharadzePhysRevAppl2016}. The same films can also be used to create superinductors in protected qubit devices \cite{KermanPRL2010, BellPRL2014, DempsterPRB2014}. Although superinductors have been realized with a chain of Josephson junctions \cite{EarnestPRL2018, LinPRL2018, PopNature2014}, a high KI nanowire is simpler to fabricate, does not suffer from spurious junction modes, and may even provide larger total inductance \cite{HazardArxiv2018}. 

In the thin film limit of superconductivity, KI scales as $\lambda_L^2/t$, where $t$ is the film thickness and $\lambda_L$ is the London penetration depth. Titanium nitride (TiN) is one of the highest known KI materials, primarily due to its large intrinsic London penetration depth, and the KI can be further increased by using ultra-thin films. Microwave resonators fabricated from thin TiN films have exhibited excellent coherence \cite{ChangAPL2013, SandbergAPL2013, VissersAPL2010}, with the highest single photon power internal quality factor exceeding $2\times10^6$ achieved using sputtering \cite{OhyaSST2013, SandbergAPL2012}. To grow wafer-scale, thin films with consistently high KI, it is important to have uniform film properties across the wafer. Atomic layer deposition (ALD) offers a conformal and repeatable film of TiN and is therefore a promising method for fabricating consistent, high KI microwave resonators. There have been successful attempts to grow TiN films using ALD \cite{CoumouIEEE2013, NahahJVST2017}, and these superconducting films have been used to study deviations from BCS theory \cite{DriessenPRL2012}.

In this work, we study high KI microwave resonators fabricated from 9 nm - 110 nm thick TiN films that are grown via ALD. Our lumped element resonator design allows for very high characteristic impedance, while maintaining a high coherence. Through a combination of the deposition method, resonator designs, and fabrication procedure we achieve high internal quality factors ($Q_i$) exceeding one million at single photon powers for resonators on thicker TiN films. On the thinnest film we achieve a characteristic impedance up to 28 k$\Omega$, while also reaching $Q_i \approx 10^5$.

\begin{figure*}
	\centering
	\includegraphics[width=6.69in]{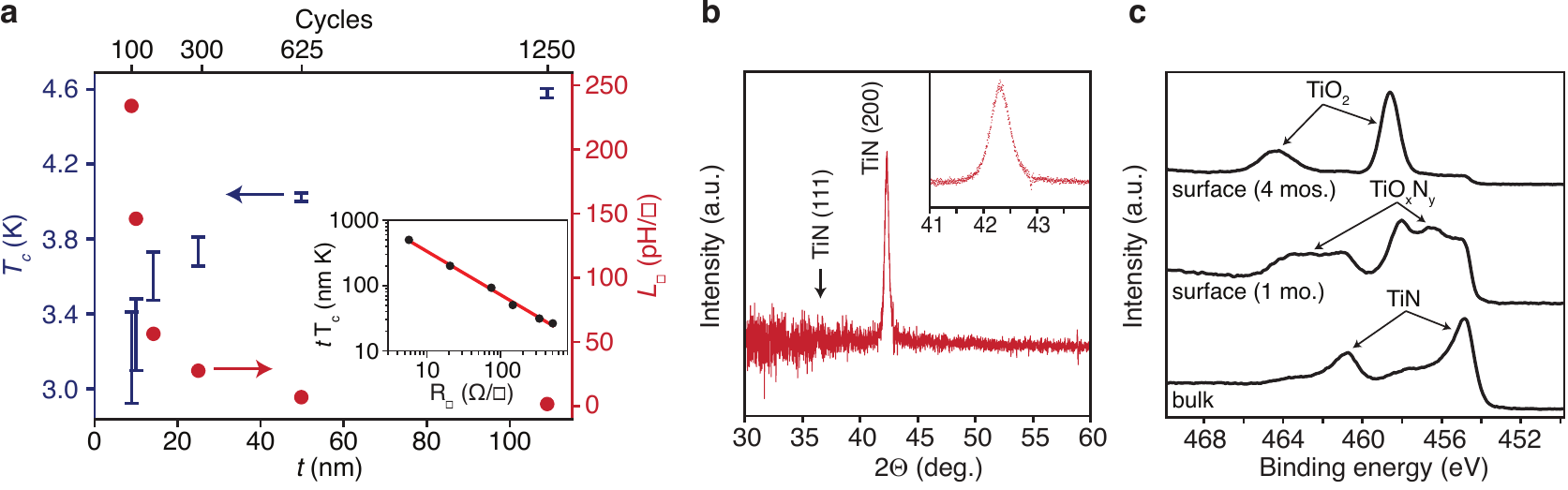}
	\caption{
		Materials characterization of TiN grown by ALD. (a) Superconducting critical temperature ($T_{c}$) and sheet inductance ($L_\square$) of TiN as a function of film thickness (\textit{t}). The bars denote the temperatures when there is a $10 \%$ and $90 \%$ reduction in resistivity from the value at 6K. The inset shows the dependence of $t T_c$ on R$_\square$, with a fit (red line) to $t T_c = A$ R$_\square^{-B}$: $A = 1554\pm 1$, $B = 0.67 \pm 0.02$. (b) $\Theta-2\Theta$ X-ray diffraction scan of TiN on Si (111). Data are background subtracted using a measurement of clean Si substrate. The increased noise at lower $2\Theta$ values is from the Si (111) Bragg diffraction peak. (c) X-ray photoelectron spectroscopy of Ti$_\mathrm{2p}$ surveys from TiN films. Bulk TiN spectra were obtained by Ar$^{+}$ ion etching into the middle of a 109 nm thick TiN film. These samples were left in air at room temperature for 1-4 months.
		\label{fig:fig1}}
\end{figure*}

In order to understand the film quality of TiN films grown by ALD, we perform materials characterization by DC electrical measurements and X-ray techniques. The superconducting TiN films used in this work are grown on hydrogen-terminated, high resistivity ($> 10~\text{k}\Omega$~cm) Si (111)-oriented substrates, which are cleaned with organic solvents, Nano-Strip, and buffered-HF immediately prior to TiN deposition. During ALD, the substrate is kept at 270 $^\circ$C, and we use Tetrakis(Dimethylamino)Titanium (TDMAT) and nitrogen gas (N$_2$) precursors. Each ALD cycle deposits approximately 0.9 \AA~of TiN, allowing for precise control over the film thickness across the entire wafer.  We find the resulting TiN surface roughness (root-mean-square) is 0.4 nm, measured by atomic force microscopy. After deposition, samples are patterned using standard optical or electron-beam lithography methods, and etched using a inductively coupled plasma with Cl$_2$, BCl$_3$, and Ar gas flow. We then record the critical temperature ($T_c$) and sheet inductance ($L_\square$) for a range of different film thicknesses (Fig. \ref{fig:fig1}a and Table \ref{tab:tab1}). The inset of Fig. \ref{fig:fig1}a shows that our films follow a universal relation that links thickness, $T_c$, and sheet resistance, which has been observed for other thin superconducting films \cite{IvryPRB2014}. Moreover, the exponent $B = 0.67\pm0.02$ we measure is similar to TiN from other work \cite{CoumouPRB2013}.

\begin{table}
	\noindent
	\caption{Properties of ALD TiN films for various thicknesses. The thickness ($t$) of each film is measured via ellipsometry, except for thicknesses marked with an asterisk, which are interpolated. The remaining parameters are determined from four wire resistance measurements. The critical temperature ($T_c$) shown is the temperature at a $50\%$ reduction in resistivity ($\rho$) from 6 K.}
	\label{tab:tab1}
	\begin{center}
		\begin{tabular} { |c|c|c|c|c| }
			\hline \hline
			Cycles & $t$ (nm) & $T_c$ (K) & $\rho$ ($\mu\Omega \cdot$cm) & $L_{\square}$ (pH/$\square$) \\
			\hline
			\hline
			100  & 8.9   & 3.01 & 449 & 234 \\
			125  & 10.7*      & 3.17 & 332 & 146 \\
			187  & 14.2  & 3.63 & 206  & 56  \\
			300  & 25.7*     & 3.76 & 186  & 28 \\
			625  & 49.8  & 4.05 & 103  & 7.1  \\
			1250 & 109.0 & 4.62 & 62  & 1.7  \\
			\hline
		\end{tabular}
	\end{center}
\end{table}

The critical temperature decreases from 4.6 K for the thickest film ($t = 109$ nm) to 3.0 K for the thinnest film ($t = 8.9$ nm), which can be attributed to disorder enhanced Coulomb repulsions \cite{EscoffierPRL2004, DriessenPRL2012}. It is worth mentioning, we were unable to produce a superconducting film with $t \approx 5$ nm (Fig. S1). Nevertheless, TiN grown by ALD thinner than $t = 8.9$ nm have been shown to go superconducting \cite{DriessenPRL2012, CoumouIEEE2013}, and it is possible that more optimization of surface preparation as well as the growth recipe could allow for thinner superconducting films by this method. 

\begin{figure*}[htbp]
	\centering
	\includegraphics[width=6.69in]{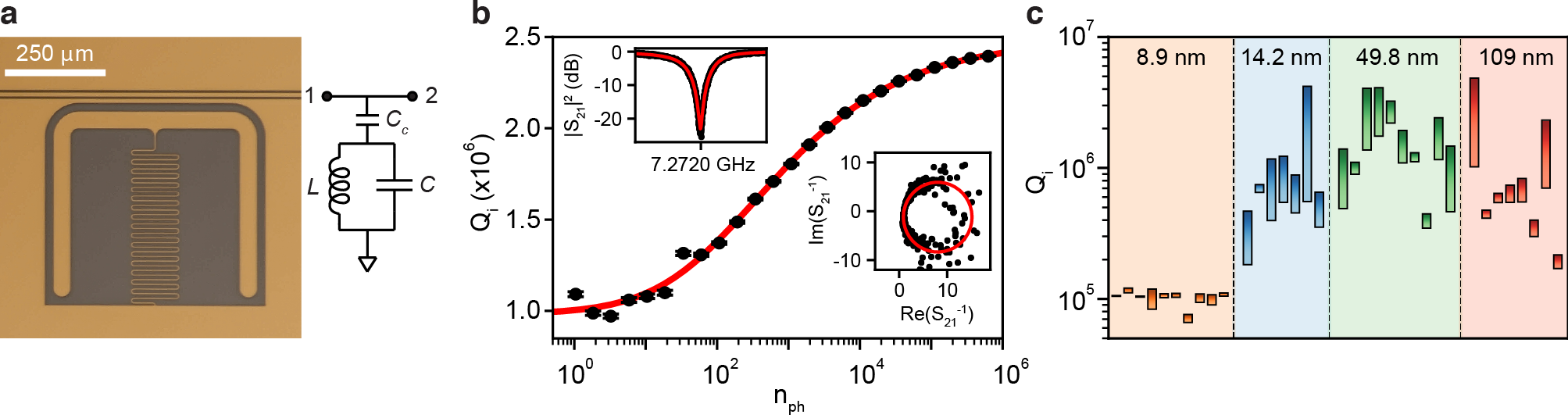}
	\caption{
		Microwave characterization of high quality factor resonators. (a) Optical micrograph of a typical resonator ($t=49.8$~nm) used in this work, which can be described by the circuit model shown on the right. We measure transmission from port 1 to port 2. (b) Power dependence of the internal quality factor for a resonator patterned on a 109 nm thick TiN film at $T = 20$~mK. The solid red line is a fit to a two-level system model that includes saturation at low and high powers. Insets show the lineshape (log magnitude, top left and inverse real and imaginary, lower right) at an average photon number $n_\mathrm{ph} \approx 3$. For this resonator $Q_c^* = 0.74\times10^5$. (c) Internal quality factors of all resonators in this study, grouped by film thickness. For a single film thickness each resonator’s internal quality factor increases with power, visualized as a bar. The bottom of each bar corresponds to single photon $Q_i$, whereas the top corresponds to the high power saturated $Q_i$ or the $Q_i$ just below bifurcation (if bifurcation was observed).
		\label{fig:fig2}}
\end{figure*}

A main feature of these films is their high kinetic inductance. We determine this quantity from $L_\square = \hbar R_\square / \pi \Delta_0$, where the sheet resistance $R_\square = \rho/t$ is measured just above $T_c$ and $\Delta_0 = 1.76 k_B T_c$ is the superconducting energy gap for TiN \cite{EscoffierPRL2004, PrachtPRB2012} predicted by BCS theory. A monotonic decrease in $T_c$ with decreasing film thickness is therefore linked to a monotonic increase in $L_\square$, which is eventually limited by the thinnest film we can grow that still superconducts. We achieve a maximum $L_\square = 234$ pH/$\square$, more than a hundredfold increase compared to the thickest film. This demonstrates the potential for TiN grown via ALD as a high KI material.

We perform X-ray diffraction to determine the dominant crystal orientation of the films. The X-ray diffraction patterns of a 109 nm thick TiN sample, chosen in order to yield the highest intensity, show a Gaussian peak at $2\Theta = 42.32^\circ \pm 0.01^\circ$ (Fig. \ref{fig:fig1}b), corresponding to (200)-oriented TiN. Furthermore, there is no detectable diffraction peak from (111)-oriented TiN at $2\Theta \approx 36.5^\circ$. We estimate a lattice parameter $a=4.26$~\AA~for the face-centered cubic TiN, which agrees with previous results using sputtering \cite{VissersAPL2010,OhyaSST2013} and first-principles calculations \cite{HaoPRB2006}. These data show that bulk TiN grown by plasma enhanced ALD is preferentially (200)-oriented on Si (111) substrates.  More material studies with synchrotron X-ray sources could help to understand the structure and ordering of ultra-thin films.

The chemical composition of the TiN is discerned by X-ray Photoelectron Spectroscopy (XPS). We find a Ti:N stoichiometric ratio of 0.96 $\pm$ 0.04, determined by comparing the Ti $2p$ and N $1s$ peak areas with binding energies of 454.9 eV and 397.2 eV, respectively. In conjunction with XPS, we use secondary ion mass spectroscopy (performed by EAG Laboratories, Inc.) to quantify various impurities. The concentrations of H, C, and O are observed to be approximately 2\%, 3\% and 1\%, respectively, throughout the bulk TiN (Fig. S2).

Since oxygen is known to affect the properties of TiN, we characterize oxide growth on the film surface by XPS measurements. Ti $2p$ spectra from the surface indicate the presence of TiO$_2$ by the 2$p_{3/2}$ peak at 458.6 eV (Fig. \ref{fig:fig1}c). The higher binding energy (2$p_{1/2}$) peaks are expected from spin-orbit coupling. Near the TiN surface, we find a thin intermediate TiO$_x$N$_y$ transition layer \cite{NahahJVST2017, SahaJAP1992}. The peak intensity of TiO$_2$ increases with aging, and we attribute this to a slow growth of a 5-8 nm thick oxide layer over a time scale of months. In addition, Ar$^+$ milling reveals that the TiO$_x$N$_y$ transition layer persists and is between the TiO$_2$ and bulk TiN. The presence and growth of an oxide layer, initiated by forming titanium oxynitride readily in atmosphere, are consistent with previous TiN film oxidation studies \cite{NahahJVST2017, LogothetidisTSF1999}. Since an amorphous oxide layer can act as a lossy dielectric, all resonators in this work are grown, patterned and mounted inside a dilution refrigerator within three days to minimize the time each film was exposed to an oxygen rich environment.

To verify that these films are low-loss at microwave frequencies, we pattern each film with a series of lumped element microwave resonators. Our design has an explicit capacitor and a meandering inductor to ground, as shown in Fig. \ref{fig:fig2}a. Each chip contains eight to ten resonators which are separated from a microwave feedline whose gap and pin width are carefully adjusted to match the 50 $\Omega$ impedance of the PCB and amplifier chain. On a single chip all resonators are designed with equal capacitance $C$ to ground. As a consequence, the resonance frequency $f_0$ is varied by adjusting the length of the inductor, while keeping the width constant at $w = 3$ $\mu$m. 

We measure resonators that vary in thickness from 8.9 nm to 109 nm and study the effect of the film thickness on the internal quality factor (see Fig. S3 for measurement setup). A typical normalized transmission spectrum at low average photon number ($n_\mathrm{ph} \approx 3$) of a resonator ($t = 109$ nm) is shown in the inset of Fig. \ref{fig:fig2}b. At the resonance frequency we observe a dip in magnitude, which is captured well by
\begin{equation}
	S_{21}^{-1} = 1 + \frac{Q_i}{Q_c^*} e^{i\phi} \frac{1}{1 + 2 i Q_i (f - f_0) / f_0},
\end{equation}
where $Q_c^*$ is the effective coupling quality factor and $\phi$ is the rotation in the complex $S_{21}^{-1}$ plane due to impedance mismatches between the resonator and the feed line \cite{MegrantAPL2012}. It is important to take this into account, especially for thin films ($t = 8.9, 14.2$ nm) where the high KI makes it difficult to avoid impedance mismatches between the feed line and PCB. Further investigation shows that $Q_i$ increases at higher microwave power (Fig. \ref{fig:fig2}b), and eventually saturates at $Q_{i, \mathrm{max}}$ for $n_\mathrm{ph} \gg 10^5$. This increase and saturation of $Q_i$ is well described by a power dependent saturation mechanism \cite{WangAPL2009, SageAPL2011}, which likely originates from two-level systems on the surface of the TiN or Si \cite{SandbergAPL2012}.

We now repeat the measurements from Fig. \ref{fig:fig2}b for four film thicknesses between $t = $ 8.9 and 109 nm and visualize the results in Fig. \ref{fig:fig2}c, where we plot $Q_i$ of four separate chips grouped by film thickness. All resonators show high internal quality factors exceeding $1 \times 10^5$. In particular, the highest $Q_i$ are obtained with the $t = 49.8$ nm sample, for which seven out of ten resonators have $Q_i (n_\mathrm{ph} = 1) > 10^6$. Since all resonators, except for those patterned on the thinnest film, show an increase in $Q_i$ with incident microwave power, $Q_i$ is most likely limited by two-level systems at $T=20$ mK and $n_\mathrm{ph} = 1$. In contrast, resonators on the thinnest film ($t=8.9$ nm) do not show this characteristic increase with power. Additionally, as the temperature is increased towards $T_c$, $Q_i$ of the thinnest resonators deviates significantly from a model derived from BCS theory (Fig. S4). This suggests $Q_i$ is limited by suppressed superconductivity instead of two-level systems. Nevertheless, these resonators still have high quality factors of $Q_i(n_\mathrm{ph}=1) \approx 1\times10^5$.

\begin{figure}[t]
	\centering
	\includegraphics[width=3.37in]{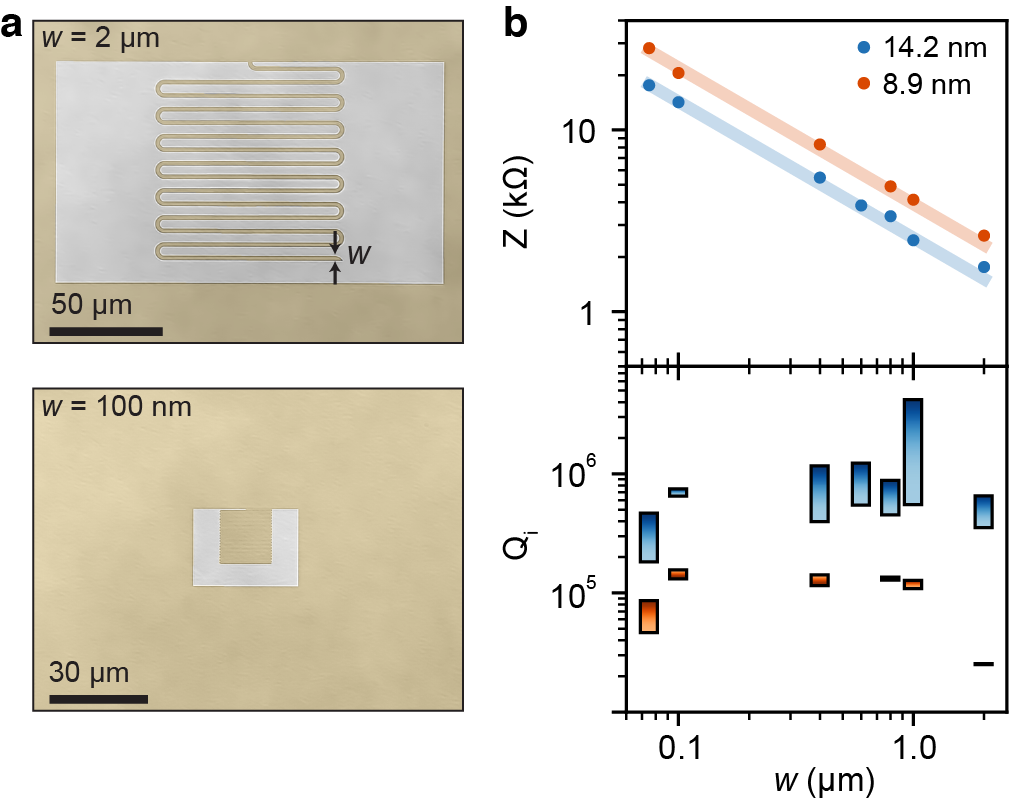}
	\caption{High $Q$, high impedance resonators. (a) Optical micrographs of high impedance TiN (false-colored yellow) microwave resonators with inductor wire width $w = 2$ $\mu$m (top) and $w = 100$~nm (bottom). Note the different scale bars. (b) Internal quality factor (top) at 20 mK and characteristic impedance (bottom) of the resonator designs shown in (a) as function of $w$. Resonators are fabricated on TiN films $t = 8.9$~nm thick (red) and $t = 14.2$~nm thick (blue). $Z$ is estimated from $2\pi f_0 (w) L_\square \ell / w$, where both $L_\square$ (see Table \ref{tab:tab1}) and $f_0(w)$ were determined experimentally. The solid lines are a guide to they eye showing a $w^{-\frac{3}{4}}$ dependence as predicted from the main text.
		\label{fig:fig3}}
\end{figure}

In order to use these resonators in hybrid systems, a high characteristic impedance is desirable since it can considerably enhance the coupling strength. We maximize the impedance of our lumped element resonator design by removing any explicit capacitor, and making the resonator a meandering wire of width $w$ and length $\ell$, as shown in Fig. \ref{fig:fig3}a. For this design the stray capacitance scales with the perimeter rather than the area of the resonator, such that the resonance frequency $f_0 \propto \left(w/\ell^3\right)^{\frac{1}{4}}$ and impedance $Z \propto \left(\ell/w^3\right)^{\frac{1}{4}}$. From this, we expect an increased impedance as we decrease $w$ from 2 $\mu$m to 75 nm, while keeping $\ell$ approximately constant.

Figure \ref{fig:fig3}b shows impedance and measured $Q_i$ as function of inductor wire width $w$ for film thickness $t = 8.9$ (red) and 14.2 nm (blue). The maximum impedance of $Z = 28$ k$\Omega$ is achieved for the thinnest film and narrowest inductor, and would result in a coupling enhancement of $\sqrt{Z/Z_0}\approx24$ times. However, even for the thickest film and widest wire the impedance $Z = 1.76$ k$\Omega$ is more than 35 times larger than a conventional 50 $\Omega$ microwave resonator. For both films, we observe no strong dependence of $Q_i$ on $w$ down to 75 nm, showing that we are able to produce high impedance resonators without having to sacrifice $Q_i$.

In conclusion, ALD grown TiN offers high quality films that can be used in detectors, hybrid systems, and protected qubits. Microwave resonators fabricated on films of thickness $t \geq 14.2$ nm showed internal quality factors exceeding $4 \times 10^5$ at single photon powers, whereas a reduced $Q_i$ was observed for the thinnest film. On the thinnest film, a modified lumped element resonator design with no explicit capacitor and footprint of only 8$\times$8~$\mu\text{m}^2$ achieved impedances up to 28 k$\Omega$. Moreover, we found no dependence of $Q_i$ on the impedance down to $w = 75$ nm. Future work includes fabrication of high impedance resonators on different high KI films, such as NbN or NbTiN, and studying even narrower inductor wires down to $w = 10$ nm, where an increased phase slip rate may affect cavity quality factors and be an interesting platform for studying the breakdown of superconductivity \cite{KuzminArxiv2018}. In this regime high KI materials can also be used in phase slip junctions \cite{MooijNatPhys2006, DeGraafNatPhys2018}.

\begin{acknowledgments} 
	The authors would like to thank S. Chakram for help with initial resonator measurements as well as A. S. Filatov for assistance with XPS studies. The authors would also like to thank P. Duda, A. Mukhortova, and A. Dixit for supporting device fabrication. We acknowledge useful discussions with Y. Lu, R. Naik, S. Hruszkewycz, and Q. Y. Tang. We thank J. Jureller for assistance with MRSEC facilities. This work was supported by the Army Research Office under Grant No. W911NF-17-C-0024. This work was partially supported by MRSEC (NSF DMR-1420709). Devices were fabricated in the Pritzker Nanofabrication Facility of the Institute for Molecular Engineering at the University of Chicago, which receives support from Soft and Hybrid Nanotechnology Experimental (SHyNE) Resource (NSF ECCS-1542205), a node of the National Science Foundation’s National Nanotechnology Coordinated Infrastructure. Work at Argonne National Lab (sample characterization) was supported by the U.S. Department of Energy, Office of Science, Basic Energy Sciences, Materials Science and Engineering Division. Use of the Center for Nanoscale Materials, an Office of Science user facility, was supported by the U.S. Department of Energy, Office of Science, Office of Basic Energy Sciences, under Contract No. DE-AC02-06CH11357.
\end{acknowledgments}

\onecolumngrid
\newpage
\appendix
\renewcommand{\thefigure}{S\arabic{figure}}
\setcounter{figure}{0}

\setcounter{equation}{0}
\numberwithin{equation}{section}
\renewcommand\theequation{\thesection\arabic{equation}}

\section{Device fabrication}
All films were grown on high resistivity ($>10~\text{k}\Omega$~cm) silicon (111) wafers (2 in. diameter) that were sourced from Virginia Semiconductor, Inc. The Si growth method was FZ, and reported thickness was $375\pm25$~$\mu$m. Surface preparation consisted of a solvent clean in an ultrasonic bath (3 min. each in NMP/IPA/Acetone/IPA), followed by a rinse in de-ionized water (DI). Immediately afterwards, the Si substrates were further cleaned using 4 minutes in Nano-Strip etch followed by DI water. The native silicon oxide was finally etched with buffered-HF for 3 minutes, and then rinsed in DI water. Samples were then blow-dried with N$_2$ and loaded into the ALD tool.

The deposition system was an Ultratech/Cambridge Fiji G2 for Plasma-Enhanced ALD. The process included a 1 hour \textit{in situ} bake at 270 $^\circ$C under vacuum, and afterwards the deposition also took place at this temperature. The titanium precursor, TDMAT, was pulsed first, followed by the nitrogen precursor, N$_2$ gas. Nitrogen gas was also used to purge the chamber between each precursor.

After removal from the ALD tool, the samples again underwent a solvent clean in an ultrasonic bath and DI rinse. We performed a dehydration bake (5 min. at 180 $^\circ$C in atmosphere) before spinning photoresist. For optical lithography we used photoresist AZ MiR 703, exposed using a Heidelberg MLA150 Direct Writer, and developed the photoresist in AZ MIF 300 followed by a rinse in DI water. For electron-beam patterning we used the resist AR-P 6200.04 and exposed using a Raith EBPG5000 Plus E-Beam Writer. These e-beam samples for high impedance resonators were developed in Amyl Acetate at 2.0 $^\circ$C, followed by a quench in IPA.

Both optical lithography and e-beam samples were etched in a Plasma-Therm inductively coupled plasma. The etch gases and flow rates were Cl$_2$ (30 sccm), BCl$_3$ (30 sccm), and Ar (10 sccm), with ICP power of 400 W and bias power of 50 W. After etching, the resist was stripped in Remover PG at 80 $^\circ$C. Samples were then diced, mounted to copper PCBs (shown in Fig. \ref{fig:exp_setup}), and wirebonded.

\section{Four point resistance measurements}
DC film characterization was performed in a Quantum Design Physical Property Measurement System (PPMS) with a base temperature of 1.8 K. All samples were kept under vacuum after RF measurements until they could be wirebonded for four-wire measurement. This was done to try to minimize any oxide growth between measurements, which might affect electrical properties of TiN films. 

\begin{figure}[hbtp]
	\centering
	\includegraphics{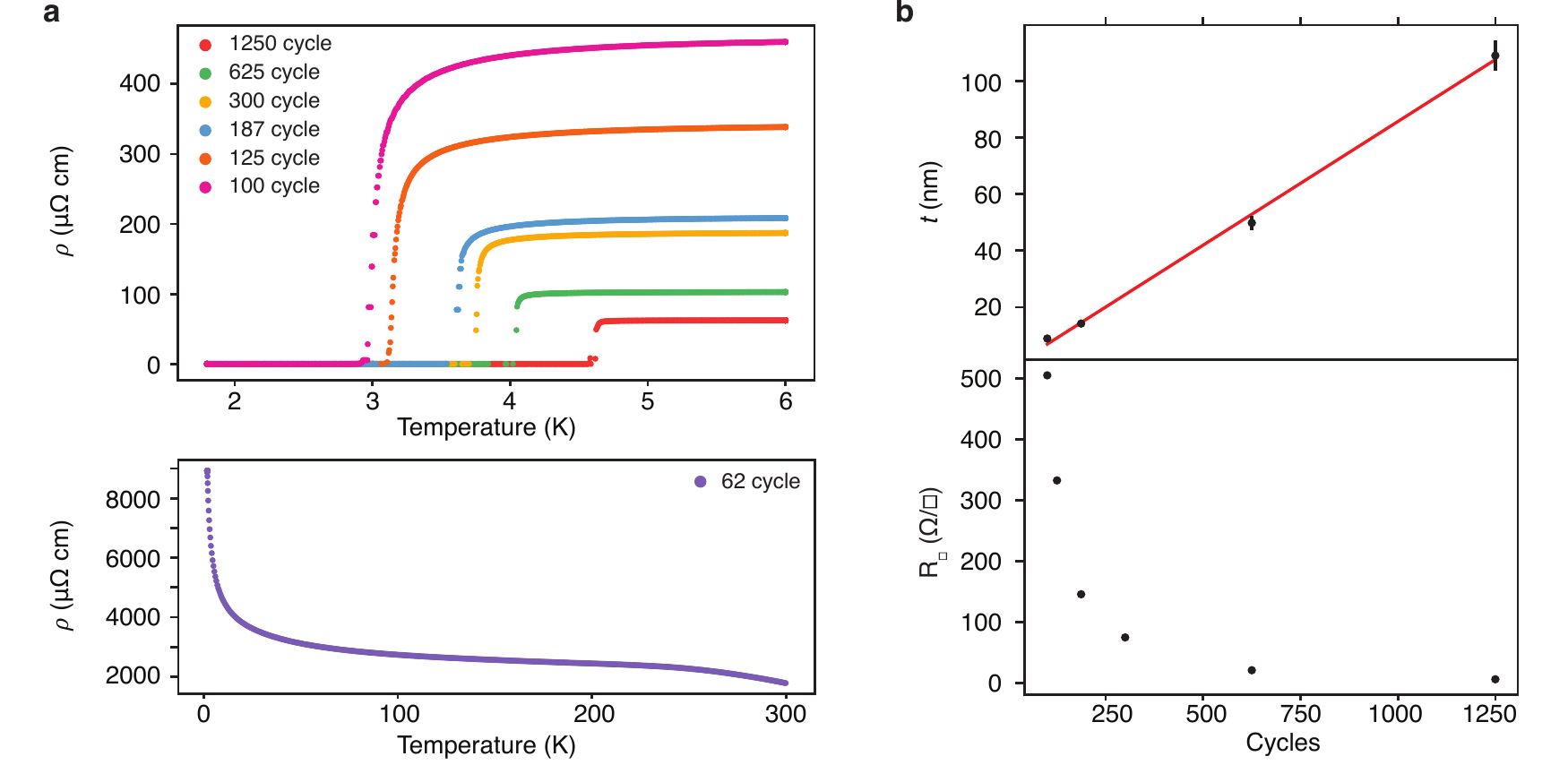}
	\caption{Four point probe measurements of TiN films (a) Resistivity as a function of temperature for various film thicknesses given by their quantities of ALD cycles. (b) Thickness (\textit{t}) and sheet resistance (R$_\square$) as a function of ALD cycles. The thickness is measured using ellipsometry and a linear fit of $t$ vs. cycles gives a growth rate of 0.9 \AA /cycle. From this growth rate we interpolate \textit{t} for 125 and 300 cycles.}
	\label{fig:rho}
\end{figure}

After letting the films thermalize at $T = 1.8$ K for one hour, we record the resistance as function of temperature during a slow temperature increase from 1.8 K to 6.0 K (rate 0.1 K/min). Results of the four point probe resistance measurements are shown in Fig. \ref{fig:rho} and were used to determine $T_c$, resistivity, sheet resistance, and sheet inductance. At $T = 1.8$ K the residual resistance was less than $1\times10^{-4} \mu\Omega$~cm for films with $t > 10$nm. The residual resistance for the 8.9 nm film was approximately $1\times 10^{-3}\mu\Omega$~cm. For three separate depositions, samples of thickness $t = 5.6$ nm showed an exponential increase in resistance with decreasing temperature (Fig. \ref{fig:rho}a) and there was no superconducting transition.

\section{SIMS of ALD TiN films and X-ray method details}
\begin{figure}[hbtp]
	\centering
	\includegraphics[width=7in]{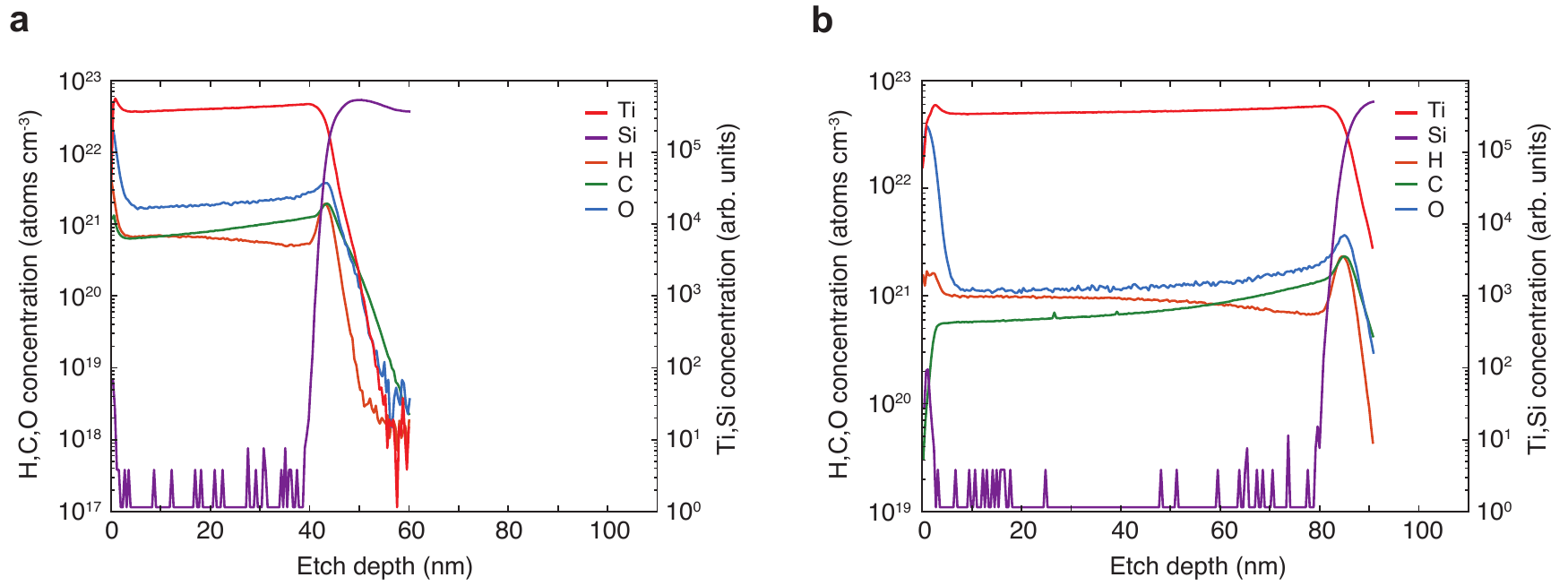}
	\caption{Secondary Ion Mass Spectroscopy (SIMS) of TiN films grown by ALD for (a) 625 cycles and (b) 1250 cycles.
		\label{fig:SIMS}}
\end{figure}

Results of secondary ion mass spectroscopy (SIMS) for careful impurity studies (hydrogen, carbon, oxygen) of 625 and 1250 cycle TiN films are shown in Fig. \ref{fig:SIMS}. The concentrations of H, C and O vary less than a few percent throughout the entire TiN film. Even though the 1250 cycle shows slightly more H relative to O contamination, its levels are similar to the 625 cycle film.

All samples used in X-ray studies were 7x7 mm wide after dicing and the TiN film was unpatterned. X-ray diffraction was performed at the Center for Nanoscale Materials at Argonne National Laboratory using a Brucker D2 Phaser XE-T. X-ray photoelectron spectroscopy (XPS) was performed at the University of Chicago, utilizing a Kratos AXIS Nova. Depth profiling was achieved by Ar$^+$ milling with a known etch rate at regular intervals and subsequently taking spectra. From XPS spectra taken after Ar$^+$ milling into the sample, we estimate there is 1-2 nm of TiO$_2$ within a month of deposition and 5-8 nm of TiO$_2$ after 4 months. Our Ti 2$p_{3/2}$ spectra results for the approximate binding energies of TiO$_2$/TiO$_x$N$_y$/TiN are 458.6/456.4/454.9 eV, respectively.

\section{Microwave measurement setup}
All characterization of the microwave resonators was done in an Oxford Triton 200 dilution refrigerator with mixing chamber temperature of 20 mK. Microwave signals traveling from room temperature to the mixing chamber plate were attenuated by approximately 100 dB, most of which comes from  cryogenic attenuators (-80 dB). The additional attenuation is accounted for by the room temperature transmission lines and stainless steel coaxial cables inside the refrigerator. 

\begin{figure}[hbtp]
	\centering
	\includegraphics[width=\textwidth]{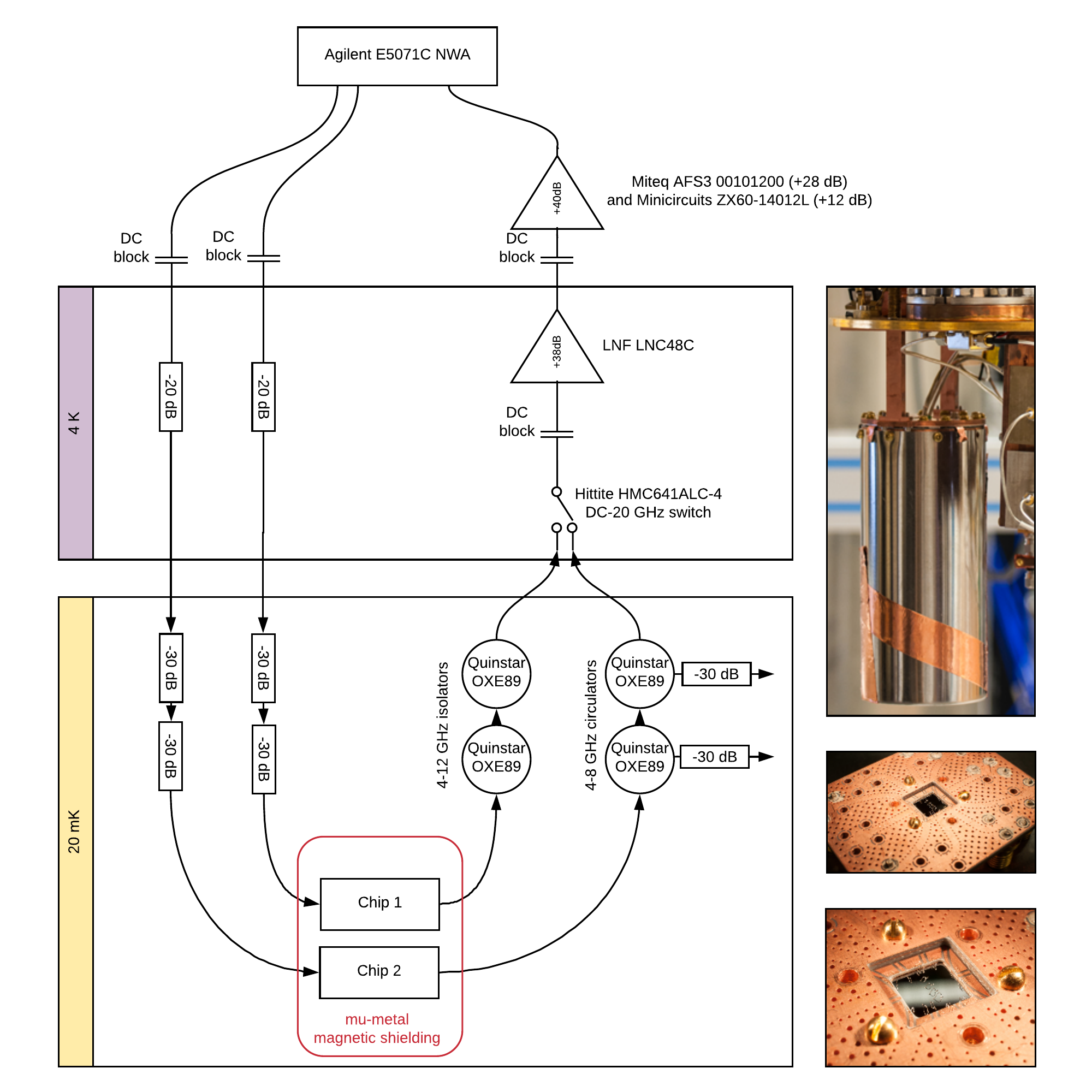}
	\caption{Schematic of the microwave measurement setup. Colored boxes highlight different temperature stages inside the refrigerator. Photographs on the right from top to bottom: Mu-metal magnetic shield mounted to the mixing chamber plate.  Printed circuit board used in this work with a chip mounted in the pocket. Zoom of a patterned 7$\times$7 mm chip with a 8.9 nm thin TiN film.}
	\label{fig:exp_setup}
\end{figure}

Each chip containing a series of resonators is mounted on a printed circuit board (See Fig. \ref{fig:exp_setup}), which is mounted inside a can made of Mu-metal magnetic shielding. The can acts as another layer of radiation shielding and protects the resonators against stray magnetic fields. The PCB is bolted to the mixing chamber plate through a copper post to provide solid thermal anchoring.

Transmitted microwaves from the resonator pass through two wideband circulators or isolators (Quinstar OXE89) after which they are amplified by a cryogenic amplifier (LNF LNC48C). A final stage of amplification at room temperature, consisting of a Miteq AFS3-00101200 and a Minicrcuits ZX60-14012L, provides 28 dB and 12 dB of gain, respectively. The full transmission is measured using a Agilent E5071C network analyzer, which records both the transmitted amplitude and phase change.

\section{Film thickness uniformity across the wafer}
\begin{figure}[hbtp]
	\centering
	\includegraphics[width=3.0in]{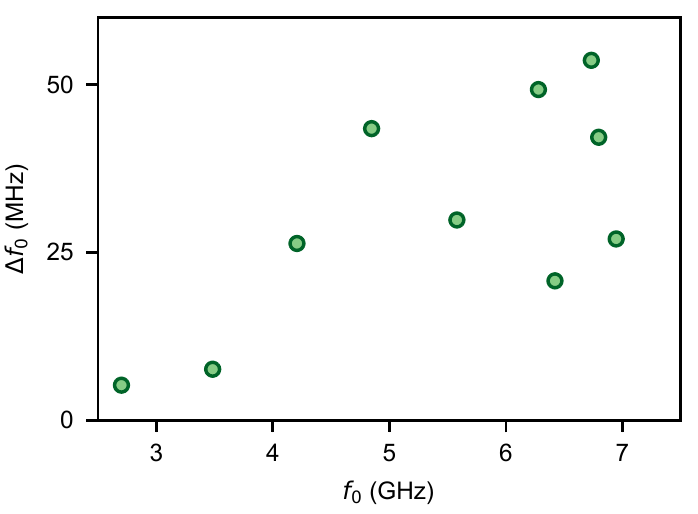}
	\caption{Resonance frequency differences between two chips from the same wafer (thickness 50 nm). We use this difference to estimate the variation in film thickness across the wafer.}
	\label{fig:FrequencyDifference50nm}
\end{figure}

To test for variations of film thickness across the wafer, we have measured resonance frequencies of two sets of identical resonators ($t = 49.8$ nm) from the same two inch diameter wafer. One of the chips was chosen from the center of the wafer, whereas the second was chosen closer to the edge. Their resonance frequencies were recorded and the result is shown in Fig. \ref{fig:FrequencyDifference50nm}. The difference in resonance frequencies $\Delta f_0$ ranges from 5 to 50 MHz, which amounts to an average fractional difference $\Delta f_0 / f_0 = 5.3 \times 10^{-3}$. If this difference were entirely due to a non-uniform film thickness across the wafer, one would expect $\Delta f_0 / f_0 = \frac{1}{2} \Delta t / t$, where $\Delta t$ is the film thickness variation across the wafer. From this we estimate $\Delta t / t \approx 1 \%$ or $\Delta t \approx 0.5$ nm. 

\section{Temperature induced frequency shift} \label{app:temp_sweeps}
\begin{figure}[hbtp]
	\centering
	\includegraphics[width=7.0in]{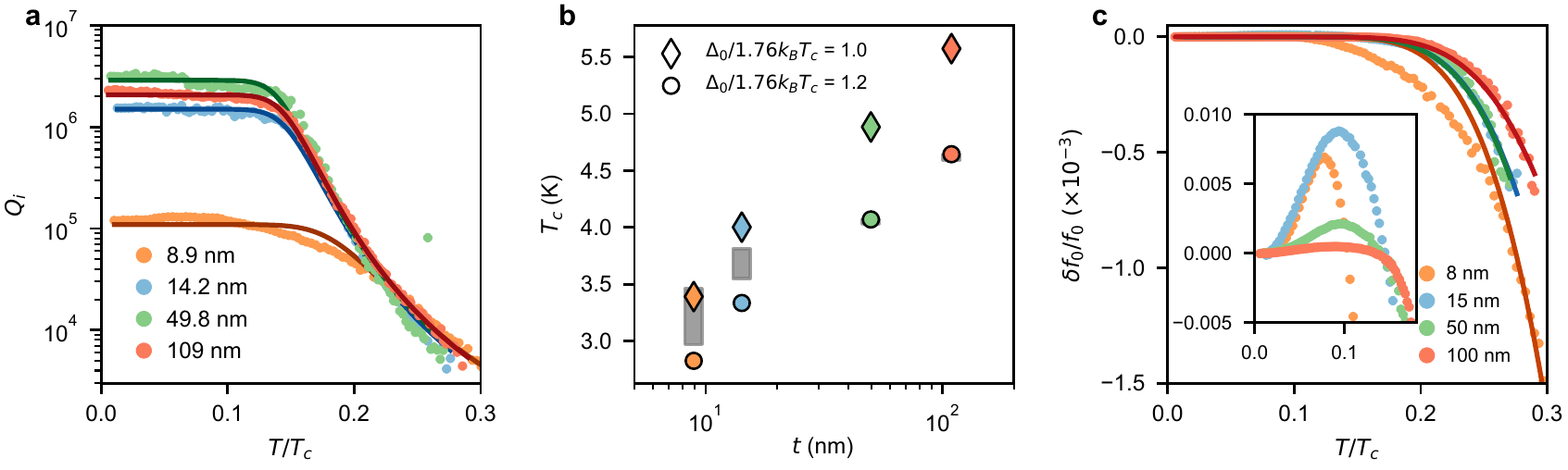}
	\caption{(a) High power internal quality factors ($n_\mathrm{ph} \approx 10^4$) as function of normalized temperature for four resonators of different film thickness. The dark, solid lines are fits to a BCS model with $T_c$ as a fit parameter. (b) Extracted $T_c$'s for two different values of $\gamma = \Delta_0 / 1.76 k_B T_c$ compared to critical temperatures obtained with four point resistance measurements (gray bars). Bottom and top of the gray bars correspond to 10\% and 90\% of the resistance normalized to the normal state resistance. (c) Fractional frequency shift as function of temperature for the same resonators as shown in (a). Solid lines are predictions from the Mattis Bardeen equations for $\sigma_2/\sigma_n$. The inset shows a zoom in of the region $T/T_c < 0.15$, highlighting an initial increase in the resonance frequency.}
	\label{fig:SupplementToFig2D}
\end{figure}

Figure \ref{fig:SupplementToFig2D}a shows a quality factor decrease as function of temperature for four different film thicknesses. The solid lines in this figure are fits to a heuristic model that includes a temperature independent upper bound $Q_{i, \mathrm{max}}$ and its functional form is given by
\begin{equation}
	Q_i^{-1}(T) = Q_{i, \mathrm{max}}^{-1} + Q_{\sigma}^{-1}(T) \label{eq:TotalQvsT}
\end{equation}
where
\begin{equation}
	Q_\sigma (T) \approx \frac{\pi}{4} \frac{e^{\Delta_0 / k_B T}}{\sinh{(h f_0/2k_B T)} K_0 \left( h f_0 / 2 k_B T \right)}. \label{eq:ZmuidzinasQsigma}
\end{equation}
$\Delta_0$ is the superconducting gap at $T = 0$, and $K_0$ is the elliptic function of the first kind. Eq. \eqref{eq:ZmuidzinasQsigma} is derived from the Mattis-Bardeen relations for the real and imaginary parts of the conductivity \cite{ZmuidzinasAnnualReview2012}. Bardeen-Cooper-Schrieffer theory predicts that $Q_\sigma$ depends on the critical temperature of the film through $\Delta_0 = 1.76 k_B T_c$. A fit of the data to Eq. \eqref{eq:TotalQvsT} with only $T_c$ and $Q_{i, \mathrm{max}}$ as free parameters yields the solid lines in Fig. \ref{fig:SupplementToFig2D}a. 

Figure \ref{fig:SupplementToFig2D}b shows extracted $T_c$'s from each of these fits as function of film thickness (diamond markers). These should be compared to the values obtained by four point resistance measurements from Table 1, which are visualized by gray bars. For each bar the height reflects the drop from the normal state resistivity (0.9 $\rho_n$) to zero resistance. We find that the fit consistently predicts higher $T_c$'s than measured by the four point resistance measurement. However, better agreement for the thickest films is obtained after rescaling $\Delta_0 = 1.76 \gamma k_B T_c$ with $\gamma=1.20$. The resulting $T_c$'s are shown as circles in Fig. \ref{fig:SupplementToFig2D}b.

Besides the internal quality factor we also record the frequency shift as function of temperature. Theoretically, the frequency shift can be described by 
\begin{equation}
	\frac{\delta f_0 (T)}{f_0(0)} = a \frac{\delta \sigma_2 (T)}{\sigma_2 (0)}, \label{eq:LinearMattisBardeenShift}
\end{equation} 
where $a = \alpha/2$, $\alpha$ is the kinetic inductance fraction and $\sigma_2$ is the imaginary part of the complex surface impedance, which is calculated by numerically integrating the Mattis-Bardeen equation for $\sigma_2/\sigma_n$ \cite{MattisBardeenPhysRev1958, TinkhamIntroToSuperconductivity, ReagorAPL2013}. Figure \ref{fig:SupplementToFig2D}c shows the measured frequency shift and prediction from BCS theory obtained by fitting the frequency shift to Eq. \eqref{eq:LinearMattisBardeenShift}. Most notably, the frequency shift of the 8.9 nm sample shows a clear deviation from the BCS prediction, whereas the thicker films fit better. Similar deviations from BCS theory have been observed for thin films of disordered TiN and NbTiN \cite{DriessenPRL2012}, and for NbN \cite{HazraArxiv2018}.

Another noteworthy feature of these data is a positive frequency shift for $T/T_c < 0.15$, as shown in the inset of the same figure. The magnitude of this frequency shift is only tens of kHz for the thinnest film and decreases with increasing film thickness. A possible candidate for this phenomenon could be two-level systems on the surface of the Si substrate or TiN \cite{GaoAPL2008, BrunoAPL2015}. Under this assumption, the frequency shift can be  written as
\begin{equation}
	\frac{\delta f_0 (T)}{f_0} = \frac{F \delta_\mathrm{TLS}}{\pi} \left[ \mathrm{Re} \Psi \left(\frac{1}{2} + \frac{1}{2\pi i}\frac{h f_0}{k_B T} \right) - \ln \left(\frac{1}{2\pi} \frac{hf_0}{k_B T} \right) \right] \label{eq:two_level_system}.
\end{equation}
Eq. \eqref{eq:two_level_system} predicts a positive frequency shift with temperature. Along with the observed power dependence of $Q_i$ shown in Fig. 2b, these data strongly suggest the presence of two-level systems in our resonators. However, Eq. \eqref{eq:two_level_system} also predicts an initial decrease in $\delta f_0$, which we do not observe. The origin of the absence is not clear.

\section{High impedance resonator design}
In this section, we derive expressions for $Z$ and $f_0$ of the lumped element, high impedance resonators in Fig. 3a. The resonance frequency is given by $f_0 = 1 / 2\pi \sqrt{L C}$, where $L = L_\square \ell / w$, $C$ is the stray capacitance from the meandering wire to ground, and $\ell$ is the wire length. The resonator's footprint can be approximated as a square with side length $d$, such that $C \approx \varepsilon A / d \approx \varepsilon \sqrt{A}$, where $\varepsilon$ is the effective dielectric constant. Therefore, the stray capacitance scales with the perimeter $d$ rather than the area $d^2$ for this resonator design. To estimate the scaling of $f_0$ with $\ell$ and $w$, we note that the actual surface area of the inductor wire is $A = \ell w$. Therefore
\begin{equation}
	f_0 = \frac{1}{2\pi} \frac{1}{\sqrt{L C}} = \frac{1}{2\pi} \frac{1}{\sqrt{L_\square \varepsilon}} \left( \frac{w}{\ell^3} \right)^{\frac{1}{4}}. \label{eq:supp_eq_f0_l_w}
\end{equation}
The scaling of the characteristic impedance $Z$ with $\ell$ and $w$ comes from the relation $Z = \sqrt{L/C} = 2\pi f_0 L$. Plugging in Eq. \eqref{eq:supp_eq_f0_l_w} yields
\begin{equation}
	Z = \sqrt{\frac{L_\square}{\varepsilon}} \left( \frac{\ell}{w^3} \right)^{\frac{1}{4}}.
\end{equation}

\end{document}